\documentclass[aps,prl,twocolumn,groupedaddress,showpacs,showkeys]{revtex4-1}
\usepackage{graphicx}
\usepackage[colorlinks=true,bookmarks=false]{hyperref}
\hypersetup{linkcolor=blue,citecolor=blue,urlcolor=blue}  
\usepackage{url}
\usepackage{natbib}
\def \t{\text}
\newcommand{\beginsupplement}{%
        \setcounter{table}{0}
        \renewcommand{\thetable}{S\arabic{table}}%
        \setcounter{figure}{0}
        \renewcommand{\thefigure}{S\arabic{figure}
				}%
     }
\usepackage{amsmath}
\usepackage{float}
\usepackage{braket}
\begin{document}

\title{Formation of a spin texture in a quantum gas coupled to a cavity}

\author{M. Landini, N. Dogra, K. Kroeger, L. Hruby, T. Donner}
\email[]{donner@phys.ethz.ch}
\author{T. Esslinger}
\homepage[]{https://www.quantumoptics.ethz.ch/}
\affiliation{Institute for Quantum Electronics, ETH Zurich, 8093 Zurich, Switzerland}
\bibliographystyle{apsrev4-1}

\date{\today}

\begin{abstract}

We observe cavity mediated spin-dependent interactions in an off-resonantly driven multi-level atomic Bose-Einstein condensate that is strongly coupled to an optical cavity. Applying a driving field with adjustable polarization, we identify the roles of the scalar and the vectorial components of the atomic polarizability tensor for single and multi-component condensates. Beyond a critical strength of the vectorial coupling, we observe a spin texture in a condensate of two internal states, providing perspectives for global dynamic gauge fields and self-consistently spin-orbit coupled gases.
\end{abstract}

\maketitle

The character of quantum many-body systems is governed by the nature of the interaction between its constituents. In combination with externally applied potentials this determines the physics of the system and the type of phenomena that can be observed. Hence, it has been an undisputed goal for experiments with quantum gases to realize new types of interactions and to expand the variety of external potentials and fields acting on the atoms. Most prominently, tunable collisional interactions between atoms with different spins led to the observation of the crossover between Bose-Einstein condensation (BEC) and Bardeen-Cooper-Schrieffer superfluidity \cite{Giorgini2008}. A more recent focus has been the engineering of long-range interactions, using spatially decaying dipolar forces \cite{Lahaye2009,Saffman2010,Baranov2012,Bohn2017} or optical cavities \cite{Ritsch2013}. Going beyond optical lattices as external potentials, synthetic gauge fields have been applied to quantum gases \cite{Dalibard2011,Goldman2014}, for example achieving spin orbit-coupling by making use of the vectorial character of atom-light interaction \cite{Lin2011}.   

Quantum gases coupled to optical cavities take a special role \cite{Ritsch2013}. The cavity field acts as a long- or global-range potential and can also be regarded as a dynamical potential back acting on all atoms, because each atom alters the properties of the field felt by all others. Off-resonantly driving a BEC coupled to an empty cavity mode gives rise to a phase transition in which a density modulation in the condensate emerges as the pump field is scattered into the cavity. This transition can be mapped to the Dicke phase transition \cite{Baumann2010,Nagy2010,Klinder2015}. The original proposal to realize the transition made use of the vectorial atom-light coupling between different internal states of a driven atomic cloud and the vacuum mode of a cavity \cite{Dimer2007}. Such an internal state Dicke transition has recently been observed with thermal atoms \cite{Zhiqiang:17}. Moreover, vectorial coupling of a BEC to a cavity has been explored in the context of spin optodynamics \cite{Brahms2010,Kohler2017}.

An open challenge for quantum gases in cavities has been to combine the control over the external degrees of freedom with vectorial atom-light coupling, thereby constructing long-range spin-spin interactions in multi-component quantum gases giving rise to spin textures and paving the way to dynamical gauge fields \cite{Zhou2009,Li2010,Guo2012,Safaei2013,Deng2014,Padhi2014,Mivehvar2015,Hung2016,Gelhausen2016,Kollath2016,Mivehvar2017}.

\begin{figure}
	\includegraphics[width=8.7cm]{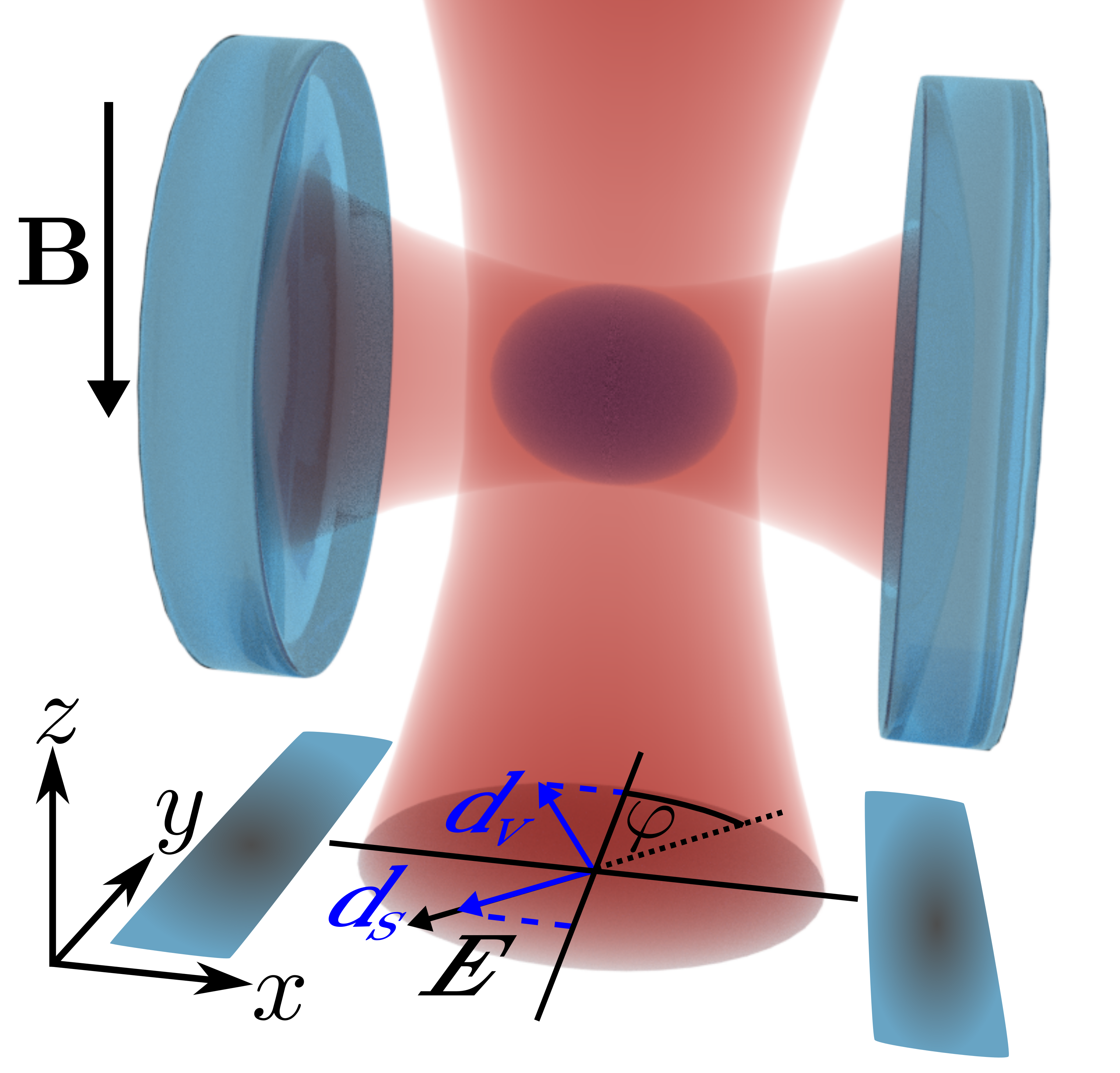}
 \caption{Pictorial illustration of the experimental system. The transverse pump, propagating along the $z$-axis, illuminates the atomic cloud which scatters light into the cavity mode, confined between the cavity mirrors. In the $x$-$y$-plane projection at the bottom we illustrate the orientation of the scalar and vectorial components of the induced dipoles $\textbf{d}_{\scriptstyle{\text{s}},\scriptstyle{\text{v}}}$ relative to the pump electric field \textbf{E}. The projection of the dipoles along the $y$-axis is relevant for the effective coupling strength to the cavity mode. An externally applied magnetic field \textbf{B} of 47 G is oriented along the negative $z$-axis.\label{fig1}}
\end{figure}

In this letter we report on the spin dependent atom-light coupling between the mode of an ultra-high finesse optical cavity and the $F$=1 total angular momentum manifold of a $^{87}$Rb BEC. The magnetic sublevels of the atoms are split by a homogenous magnetic field and the gas is prepared either in one of the three magnetic sublevels, or in a balanced mixture of the $m_F=+1$ and $m_F=-1$ sublevels. The atoms are exposed to an off-resonant laser field in standing wave configuration, which crosses the cavity perpendicularly and which we refer to as transverse pump, see Fig.~\ref{fig1}. This pump field induces rotating dipoles in the multi-level atomic gas, which can be decomposed into two orthogonal linear components that oscillates out of phase with each other. These dipoles can radiate into the cavity mode and reduce the potential energy of the system, as a consequence of the dynamic Stark shift experienced by the atoms. Simultaneously, the scattering of photons promotes the atoms to higher momentum states, setting a cost in kinetic energy for this process. For atoms prepared in only one of the magnetic sublevels the following scenario is therefore expected. At a critical pump power, the competition between kinetic and potential energy leads to a phase transition of the system from the normal state, where scattering from the homogeneous gas is suppressed due to destructive interference, to a self-organized state, in which the atomic gas is building up a density modulation and scatters photons into the cavity \cite{Dimer2007}. The intra-cavity photon number, experimentally accessible via the light leaking from the cavity, can conveniently be used as an order parameter for the phase transition.

The system can be described as an ensemble of induced dipoles. For the relevant wavelength range employed in the rest of the letter, the atomic dipole operator is given by \cite{LeKien2013}:
\begin{equation}
\hat{\textbf{d}}=\hat{\textbf{d}}_s+i\hat{\textbf{d}}_v=-\frac{\alpha_{\scriptstyle{\text{s}}}}{2} \hat{\textbf{E}}+i\frac{\alpha_{\scriptstyle{\text{v}}}}{4 F} \hat{\textbf{F}}\times\hat{\textbf{E}}
\label{d}
\end{equation}
where $\alpha_{\scriptstyle{\text{s}},\scriptstyle{\text{v}}}$ are the scalar and vectorial components of the polarizability tensor, $\hat{\textbf{F}}$ is the atomic spin vector in the ground state manifold, $F(F-1)$ is the eigenvalue of $\hat{\textbf{F}}^2$, and $\hat{\textbf{E}}$ is the electric field at the position of the atom.
We define the polarization axis of the transverse pump by the angle $\varphi$ relative to the $y$-axis. It controls the direction of oscillation of the real and imaginary components of $\hat{\textbf{d}}$, see Eq.~(\ref{d}) and Fig.~\ref{fig1}. The scalar component $\hat{\textbf{d}}_s$ oscillates in phase with the pump field and in the direction of the pump polarization, while the vectorial component $\hat{\textbf{d}}_v$ oscillates $90^\circ$ out of phase and in the direction orthogonal to both the polarization of the pump and the atomic spin. Their strengths are proportional to $\alpha_{\scriptstyle{\text{s}}}$ and $\alpha_{\scriptstyle{\text{v}}}$, respectively. The cavity cannot accept radiation components whose electric field oscillates in the direction of the cavity axis, i.e. along the $x$-axis. The coupling to the cavity is hence proportional to the projection of the dipole to the $y$-$z$-plane as indicated in Fig.~\ref{fig1}.  

Our experiment starts with a BEC of approximately 4$\times$10$^4$ $^{87}$Rb atoms trapped at the position of a fundamental mode of a high-finesse optical cavity \cite{Baumann2010}. 
The wavelength of the transverse pump $\lambda$ is 784.7 nm, and the orientation of its linear polarization is controlled through motorized waveplates \cite{Details2018}. The Zeeman splitting of the magnetic levels is set to 30 MHz to prevent spin changing processes. 

\begin{figure}
	\includegraphics[width=8.7cm]{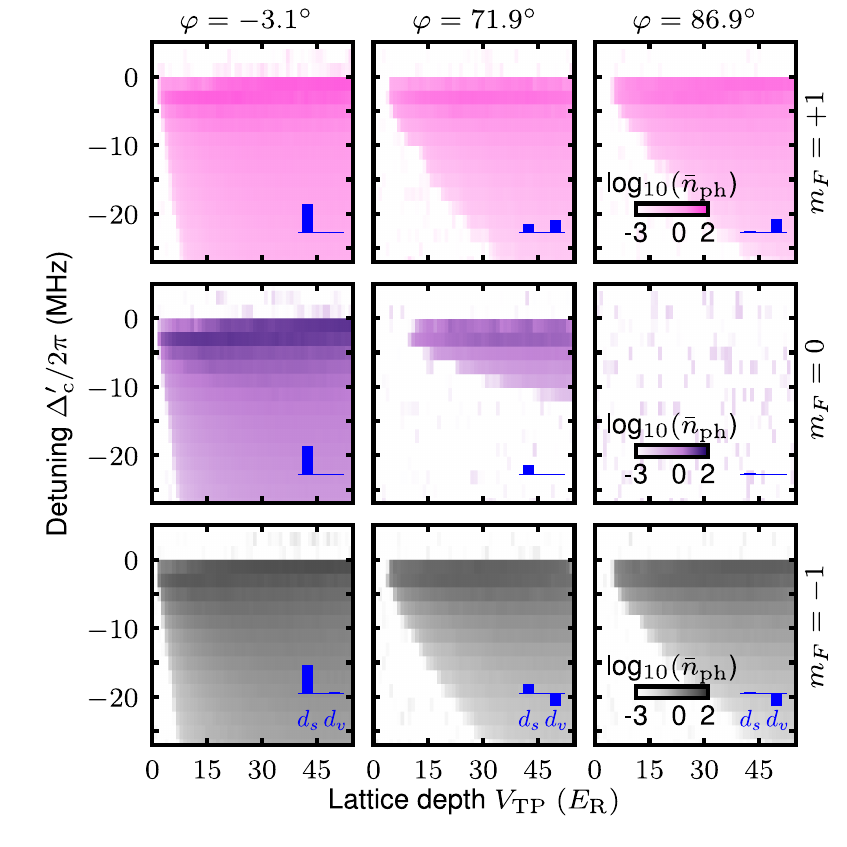}
 \caption{Measured phase diagrams as a function of lattice depth $V_{\scriptstyle{\text{TP}}}$, for the three Zeeman states ($m_F=0,\pm1$) and for three polarization angles $\varphi$. The $m_F=0$ state progressively looses coupling for increasing $\varphi$ since it couples only to the scalar component of the polarizability tensor. The other two states couple via both scalar and vectorial components, and do not differ in relative strength. For clarity of exposition, we plot against the effective detuning $\Delta_{\scriptstyle{\text{c}}}'$, corrected for cavity birefringence \cite{Details2018}.\label{fig2}}
\end{figure}

In an initial experiment, we vary the orientation of the induced atomic dipoles via the polarization angle $\varphi$ (see Fig.~\ref{fig1}) and explore the phase diagram of our system, see Fig.~\ref{fig2}.
 In the experimental sequence, we ramp up the transverse pump lattice depth $V_{\scriptstyle{\text{TP}}}$ over 60 ms at fixed detuning $\Delta_{\scriptstyle{\text{c}}}=\omega_{\scriptstyle{\text{p}}}-\omega_{\scriptstyle{\text{c}}}$ between the pump frequency $\omega_{\scriptstyle{\text{p}}}$ and the cavity resonance frequency $\omega_{\scriptstyle{\text{c}}}$.  Due to the presence of two slightly detuned cavity modes, we define their mean detuning as effective detuning $\Delta_{\scriptstyle{\text{c}}}'$ \cite{Details2018}. We constantly monitor the light field leaking from the cavity using a heterodyne detection system \cite{Baumann2010}. From this measurement we infer the number of intra-cavity photons $\bar{n}_{\scriptstyle{\text{ph}}}$, which becomes macroscopic when entering a self-organized phase.

When the polarization of the pump is oriented perpendicular to the cavity axis ($\varphi\simeq0^\circ$), only the scalar dipole component can emit into the cavity, leading to a spin-independent phase diagram. As the polarization is turned towards the cavity axis ($\varphi\simeq90^\circ$), the response of the system is progressively dominated by the spin-dependent vectorial component of the dipoles, see Fig.~\ref{fig2}. 
We introduce a two modes description for the Hilbert space of the BEC \cite{Details2018}. For zero lattice depth $V_{TP}$, the energy separation between the modes is given by 2 $E_{\scriptstyle{\text{R}}}$, where $E_{\scriptstyle{\text{R}}}=h\times$3.73 kHz is the recoil energy corresponding to the momentum imparted by a single photon. 
The many-body Hamiltonian, for atoms in a single spin component, reads
\begin{eqnarray}
 \hat{H}_{m_F}&=&-\hbar\widetilde{\Delta}_{\scriptstyle{\text{c}}} \hat{a}^\dagger \hat{a}+\hbar\omega_0 \hat{J}_{z,m_F}+\frac{\hbar}{\sqrt{N}}\left( \lambda_{\scriptstyle{\text{s}}} \cos\varphi(\hat{a}+\hat{a}^\dagger)+\right. \nonumber\\
&&\left. +i\lambda_{\scriptstyle{\text{v}}} m_F\sin\varphi(\hat{a}^\dagger-\hat{a}) \right)\hat{J}_{x,m_F},
\end{eqnarray}
 where $\hat{\textbf{J}}_{m_F}$ is the angular momentum operator for the two modes Hilbert space of each magnetic sublevel, $\widetilde{\Delta}_{\scriptstyle{\text{c}}}$ is the detuning from cavity resonance including dispersive shift, $\hbar\omega_0$ is the bare energy of the excited mode, $\hbar\lambda_{\scriptstyle{\text{s}},\scriptstyle{\text{v}}}$ denote scalar and vectorial coupling energies to the cavity mode, $\hat{a}$ is the destruction operator for a cavity photon and $N$ is the number of atoms \cite{Details2018}. 
 The Hamiltonian $\hat{H}_{m_F}$ can be rewritten in the form
\begin{eqnarray}
 \hat{H}_{m_F}&=&-\hbar\widetilde{\Delta}_{\scriptstyle{\text{c}}} \hat{a}^\dagger \hat{a}+\hbar\omega_0 \hat{J}_{z,m_F}+\nonumber\\
&&+\frac{\hbar}{\sqrt{N}}(\lambda_{m_F}^* \hat{a}+\lambda_{m_F} \hat{a}^\dagger)\hat{J}_{x,m_F},
\label{Hmf}
\end{eqnarray}
where we have introduced the complex coupling $\lambda_{m_F}=|\lambda_{m_F}| e^{i\phi_{m_F}}$ with
\begin{eqnarray}
 &&|\lambda_{m_F}|=\sqrt{\lambda_s^2 \cos^2\varphi+\lambda_v^2 m_F^2\sin^2\varphi },\label{lc}\\
&&\phi_{m_F}=\arctan\left(\frac{\lambda_v m_F}{\lambda_s}\tan\varphi\right).\label{phi}
\end{eqnarray}
Using the phase transformation $\hat{a}\rightarrow\hat{a}e^{-i\phi_{m_F}}$, Eq.~(\ref{Hmf}) takes the form of the Dicke model \cite{PhysRev.93.99}. 
The critical point for the lattice depth $V_{\scriptstyle{\text{TP}}}$ is
\begin{equation}
V_{\scriptstyle{\text{TP}}}^{\scriptstyle{\text{c}}}=\frac{\hbar\omega_0(\widetilde{\Delta}_{\scriptstyle{\text{c}}}^2+\kappa^2)}{ 16N M_0^2\widetilde{\Delta}_{\scriptstyle{\text{c}}} U_0}\frac{1}{ \cos^2\varphi+\left(\frac{\alpha_{\scriptstyle{\text{v}}} m_F}{2F\alpha_{\scriptstyle{\text{s}}}}\right)^2 \sin^2\varphi },
\label{vc}
\end{equation}
where we have introduced the maximum dispersive shift of the cavity resonance induced by a single atom $U_0$, the cavity loss rate $\kappa$ and the overlap integral $M_0$. We incorporate interactions due to atomic s-wave collisions in the value of $\omega_0$ \cite{Details2018}. 

\begin{figure}
	\includegraphics[width=8.7cm]{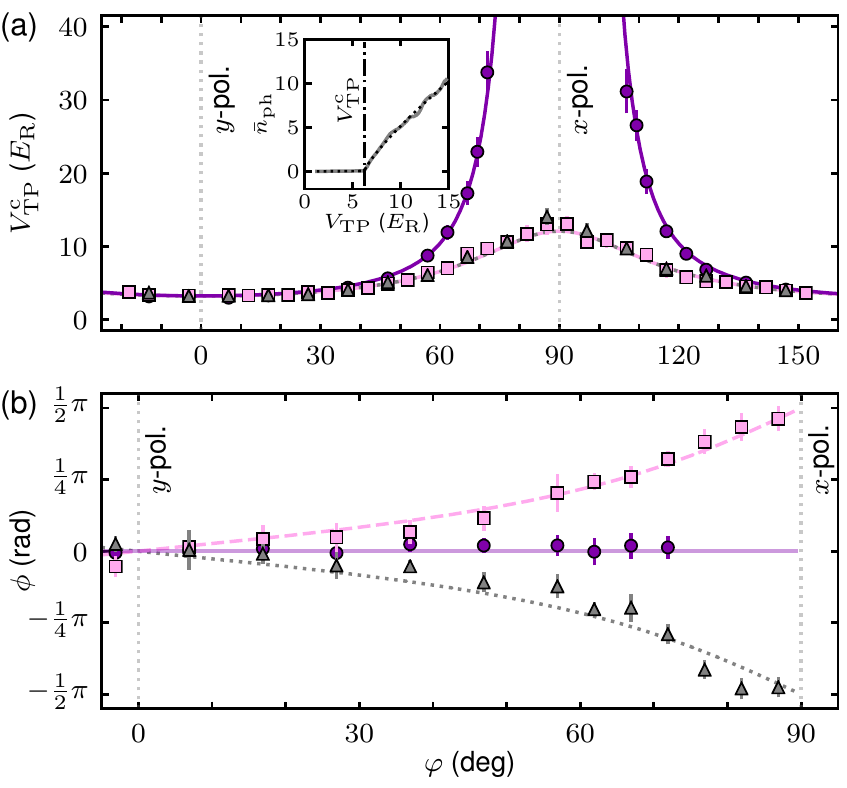}
 \caption{(a) Threshold lattice depth $V_{\scriptstyle{\text{TP}}}^{\scriptstyle{\text{c}}}$ for atoms in state $m_F$=0 (circles), +1 (squares) and -1 (triangles) for a fixed cavity detuning $\Delta_{\scriptstyle{\text{c}}}'$ of -2$\pi\times$11.36(19) MHz as a function of pump polarization angle $\varphi$. A typical measurement of the critical pump strength is shown in the inset. The output light field is analyzed to extract a threshold value for the lattice depth. (b) Detected phase $\phi_{1(-1)}$ (squares and triangles, respectively) of the intra-cavity light field for detuning $\Delta_{\scriptstyle{\text{c}}}'=-2\pi\times$8.79(18) MHz for the spin components $m_F=\pm1$, relative to the phase $\phi_0$ (circles) of the spin component $m_F=0$. The lines are our theoretical expectations, taking an experimental offset in the $\varphi$-axis and a global scaling of the $y$-axis in (a) as free parameters. We fix the value of $\alpha_{\scriptstyle{\text{v}}}/\alpha_{\scriptstyle{\text{s}}}=0.928$, as expected from theory. We are not able to measure $\phi_0$ for polarization angles close to $\pi/2$, since the threshold for self-organization diverges around that angle. Due to the symmetry breaking mechanism at the phase transition \cite{Baumann2011}, all of the measured phases are defined modulo $\pi$. Error bars are the sum of statistical and systematic contributions \cite{Details2018}.\label{fig3}}
\end{figure}

We measure the critical lattice depth $V_{\scriptstyle{\text{TP}}}^{\scriptstyle{\text{c}}}$ for self-organization by ramping up the power at fixed detuning $\Delta_{\scriptstyle{\text{c}}}'$ and repeat this measurement for different $\varphi$. Fig.~\ref{fig3}(a) depicts the angle and spin state dependent threshold for self-organization.  
 We measure the same threshold value for $m_F=\pm1$, while a higher value is obtained for $m_F$=0. Eq.~(\ref{vc}) closely reproduces the observed functional form.

The phase $\phi_{m_F}$ originates from the time delay of the field emitted by the rotating atomic dipoles and corresponds to the relative phase shift between the pump and cavity field in the organized phase.
We measure the phase of $\langle\hat{a}\rangle$ with the heterodyne detection setup \cite{Details2018}. The results are presented in Fig.~\ref{fig3}(b) where we plot the phase relative to $\phi_0$. The phase difference between $m_F=0$ and $m_F=1$ evolves from 0 to $\pi/2$ when rotating the polarization from $0^\circ$ to $\varphi=90^\circ$, while throughout the plot $\phi_{-1}=-\phi_1$ as predicted from Eq.~(\ref{phi}).

\begin{figure}
	\includegraphics[width=8.7cm]{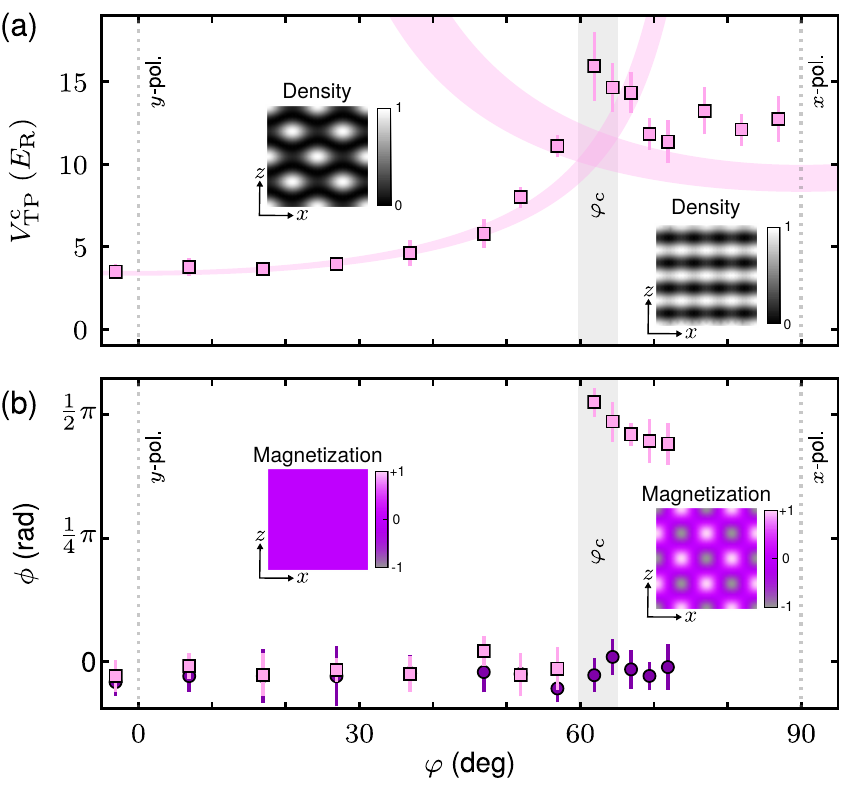}
 \caption{(a) Threshold lattice depth $V_{\scriptstyle{\text{TP}}}^{\scriptstyle{\text{c}}}$ for self-organization of the $m_F=\pm 1$ spin mixture as a function of transverse pump polarization angle $\varphi$ for a detuning $\Delta_{\scriptstyle{\text{c}}}'=-2\pi\times10.71(19)$ MHz. The shaded regions are our theoretical expectation from Eq.~(\ref{sc},\ref{vcmix}), taking into account an imperfect overlap between the two spin components. (b) Phase of the light field emitted by the spin mixture in the organized phase as a function of the pump polarization (squares). A distinct step is recognizable at a critical polarization angle $\varphi_{\scriptstyle{\text{c}}}$ of around $60^\circ$, the shaded vertical bar corresponds to our theoretical expectation for $\varphi_{\scriptstyle{\text{c}}}$ when accounting for collisional interactions as well as imperfect overlap due to spurious magnetic gradients. The purple circles are the reference phase for organization in $m_F = 0$. Error bars are the sum of statistical and systematic contributions. Insets show schematics of density and magnetization patterns in the organized states. \cite{Details2018}.\label{fig4}}
\end{figure}

We use the framework of scalar and vectorial coupling to describe the self-organization of the BEC in the more general case of a mixture of spin states. Since spin changing processes are suppressed, we can consider the atom number in each spin state $N_{m_F}$ as a constant. The Hamiltonian describing self-organization of the spin mixture reads
\begin{eqnarray}
 \hat{H}_{\scriptstyle{\text{mix}}}&=&-\hbar\widetilde{\Delta}_{\scriptstyle{\text{c}}} \hat{a}^\dagger \hat{a}+\hbar\omega_0 \sum_{m_F} \hat{J}_{z,m_F}+\nonumber\\
&&+\sum_{m_F}\frac{\hbar}{\sqrt{N_{m_F}}}(\lambda_{m_F}^* \hat{a}+\lambda_{m_F} \hat{a}^\dagger)\hat{J}_{x,m_F}.\label{mix}
\end{eqnarray} 
The light fields scattered by the individual spin components interfere, leading to competition between different self-organized states. To explore such competition we prepare a balanced mixture in the $m_F=+1$ and $m_F=-1$ Zeeman states, $N_1=N_{-1}=N/2$ leading to couplings $\lambda_{\pm1}$ of the same magnitude.  
 In analogy to the previous measurements, we ramp up the pump power at a constant detuning and measure threshold lattice depth and phase shift for different polarization of the pump, see Fig.~\ref{fig4}. 

The threshold lattice depth $V_{\scriptstyle{\text{TP}}}^{\scriptstyle{\text{c}}}$ peaks at a critical value of the angle $\varphi_{\scriptstyle{\text{c}}}$, for which a distinctive step of about $\pi/2$ is observed in the phase of the light. Both features signal where the dipole coupling switches from scalar to vectorial. 

To interpret our experimental findings, we analyze Hamiltonian Eq.~(\ref{mix}) at the mean-field level \cite{Details2018}. Below threshold, $\langle \hat{J}_{x,m_F}\rangle=0$ for all $m_F$ states and $\langle\hat{a}\rangle=0$. Above threshold, there are two possible steady-states for the system, distinguished by the relative sign of $\langle \hat{J}_{x,1}\rangle$ and $\langle \hat{J}_{x,-1}\rangle$. The value of the polarization angle $\varphi$ decides which steady-state is energetically favored.  
The transition from the normal phase to each of the two organized states has the same behavior as in the Dicke model, but gives different organization patterns, see insets in Fig.~\ref{fig4}. A change of sign in $\langle\hat{J}_{x,m_F}\rangle$ produces a shift in the density maxima of the organized pattern by $\lambda/2$. If $\langle\hat{J}_{x,\pm 1}\rangle$ have the same sign, the density maxima of the two components coincide, corresponding to a density modulated phase with zero magnetization. If the signs are opposite, the magnetization is modulated with a period $\lambda$ and the atomic density with a period $\lambda/2$, thus forming a spin texture due to cavity mediated spin-dependent interactions between the atoms. The organized state with higher energy corresponds to a suboptimal state, producing metastable configurations of the system. 

Neglecting cavity dissipation, we can analytically obtain the critical lattice depth for the density modulation as
\begin{eqnarray}
 V_{\scriptstyle{\text{TP}}}^{\scriptstyle{\text{c}},\scriptstyle{\text{d}}}&=&\frac{\hbar\omega_0\widetilde{\Delta}_{\scriptstyle{\text{c}}}}{4 N M_0^2U_0}\frac{1}{\cos^2\varphi}\label{sc},
\end{eqnarray}
and for the spin pattern as
\begin{eqnarray}
 V_{\scriptstyle{\text{TP}}}^{\scriptstyle{\text{c}},\scriptstyle{\text{s}}}&=&\frac{\hbar\omega_0\widetilde{\Delta}_{\scriptstyle{\text{c}}}}{ N M_0^2U_0}\left(\frac{\alpha_{\scriptstyle{\text{s}}}}{\alpha_{\scriptstyle{\text{v}}}\sin\varphi}\right)^2.\label{vcmix}
\end{eqnarray}
 For the density modulation, the phase of the cavity field is always $0$ mod$(\pi)$, while it is $\pi/2$ mod$(\pi)$ for the spin texture. Eq.~(\ref{sc}) leads to purely scalar coupling. In this case the system is behaving exactly as if the atoms were all in the $m_F=0$ state (effective pairing). In Eq.~(\ref{vcmix}) instead the scalar coupling is suppressed completely. Assuming that the two critical points are different, the state corresponding to the lower $V_{\scriptstyle{\text{TP}}}^{\scriptstyle{\text{c}}}$ will become macroscopically occupied. The two couplings are equal for a critical angle of 
\begin{eqnarray}
 \varphi_{\scriptstyle{\text{c}}}=\arctan\left(\frac{2\alpha_{\scriptstyle{\text{s}}}}{\alpha_{\scriptstyle{\text{v}}}}\right)\label{pc}.
\end{eqnarray}
 
Fig.~\ref{fig4} shows our measurements and the predictions of the mean-field model. Eq.~(\ref{pc}) predicts the value $\varphi_{\scriptstyle{\text{c}}}=65.1^\circ$, which differs by a few degrees with respect to the position of the $\pi/2$ phase step in Fig.~\ref{fig4}(b). Taking into account collisional interactions, our prediction is compatible with the observed value of $\varphi_{\scriptstyle{\text{c}}}$. However, our model underestimates the threshold for organization in the spin pattern. This can be caused by additional effects due to spurious magnetic gradients in our system as well as by cavity dissipation. 

We showed that the effects of vectorial polarizability are strong enough to drive new phases with magnetic ordering. We expect that dissipative effects are enhanced for competing organization patterns, which constitutes a promising future research direction. In addition, extending the coupling scheme to both polarization modes of the cavity, magnetic configurations with higher symmetries can be investigated \cite{Moodie2018}. The generation of global dynamical gauge fields as well as self-consistent spin-orbit coupled gases is a natural extension of the results reported \cite{Deng2014,Padhi2014,Mivehvar2015,Kollath2016}.   

\begin{acknowledgments}
 We thank S. Parkins and F. Mivehvar for discussions. We acknowledge funding from SNF for NCCR QSIT, and SBFI support for the Horizon2020 project QUIC.
\end{acknowledgments}
\bibliographystyle{apsrev4-1}
%
\newpage
\beginsupplement
\begin{widetext}

\section{Supplementary informations}

\section{Experimental details\label{Ed}}
\subsection{Sample preparation}

Following a radio-frequency evaporation in a magnetic quadrupole trap, the atoms initially end up in the $\ket{F=1,m_F=-1}$ hyperfine state where $F$ and $m_F$ represent the total angular momentum and the corresponding magnetic quantum number. To successively prepare the cloud in different spin states $m_F=0,+1,-1$, we perform Landau-Zener sweeps starting from $m_F=-1$. The sweeps are performed while the cloud is optically transported \cite{Brennecke2007s} from the magnetic quadrupole trap to the center of the cavity mode. The atomic cloud is loaded in a crossed dipole trap with trapping frequencies of $\omega_{x,y,z}/2\pi = [201(9),35(2),174(3)]$ Hz at the center of the cavity mode. 

For the threshold and phase diagram measurement on a single spin cloud, we prepare a BEC of $3.79(14) \times 10^4$ $^{87}$Rb atoms in different $m_F$ states. For the relative phase measurement of the scattered light by different spin states, we start with $5.7(4)\times 10^4$ atoms in $m_F=+1$. After changing the spin state of the atoms, we can either obtain $4.5(7)\times 10^4$ atoms in $m_F=-1$ (and the remaining atoms in $m_F=0$ and $m_F=+1$ spin state) or $3.0(3)\times 10^4$ atoms in $m_F=0$ (which is purified after spin change). To perform the measurements on a spin mixture, we start with $4.0(2)\times 10^4$ atoms in $m_F=0$ state and alter the spin state to obtain $2.1(3) \times 10^4$ atoms in $m_F=+1$ and $2.1(4) \times 10^4$ atoms in $m_F=-1$.\\

\subsection{Transverse pump characterization}
The lattice depth of the transverse pump is calibrated via Raman-Nath diffraction \cite{Morsch2006s}. We carefully characterized the polarization of the transverse pump by measuring it just after exiting the window of the vacuum chamber. The incident transverse pump beam is retro-reflected to form an optical standing wave. The polarization of the incident laser beam is controlled by a pair of half- and quarter-waveplates mounted on a motorized waveplate mount. The polarization of the retro-reflected beam is matched with the incident beam by placing a pair of half- and quarter-waveplates, and analyzed with a polarizing beam splitter before the retro-reflector. The polarization of the retro-reflected beam is identical to the polarization of the incident beam when all the light reaches the retro-reflector after the polarizing beam splitter. This analysis is carried out for every polarization of the transverse pump (set by the motorized waveplates) by changing the orientation of the two waveplates just before the polarizing beam splitter. In order to perform self-organization, we increase the lattice depth of the transverse pump via a smooth ramp (S-ramp) of the following form: $V(t) = V_0[3(t/t_0)^2 - 2(t/t_0)^3 ]$ where $V_0$ is the final lattice depth of the transverse pump lattice and $t_0$ is the duration of the ramp.\\

\subsection{Changing the spin state of the atoms}
In order to change the spin state of the atoms, we apply a resonant radio frequency (rf) pulse. For technical reasons we can only apply radio frequency signal below 1 MHz. We therefore reduce the magnetic field from 47 G to 114 mG via a smooth S-ramp in 50 ms (Fig. \ref{figSI}). This magnetic field corresponds to a Zeeman splitting of 80 kHz between two neighboring spin states. When measuring the relative phase of the scattered light between different spin states, we start with atoms in $m_F=+1$ and by the appropriate choice of the duration of the rf-pulse we can prepare atoms in either $m_F=-1$ or $m_F=0$. For a duration of 30 $\mathrm{\mu s}$ ($\pi-$pulse), we can prepare a state with a population of 78(11)\% in $m_F=-1$, 22(9)\% in $m_F=0$ and 0.3(1.8)\% in $m_F=+1$. Inefficiency of the $\pi$-pulse to obtain pure $m_F=-1$ state arises likely from slow drifts of the magnetic field from one cycle of the experiment to another. By choosing the duration of the rf-pulse to be 17 $\mathrm{\mu s}$ ($\pi/2-$pulse), we prepare the state $\frac{1}{2}(\ket{+1} + \sqrt{2}e^{i\phi_0}\ket{0} + e^{i\phi_{-1}}\ket{-1})$, which has the largest fraction of $m_F=0$ that one can achieve while starting from $m_F=+1$ state for negligible quadratic Zeeman shift. The phases $\phi_0$ and $\phi_1$ are oscillating in time due to the finite magnetic field. By applying a strong magnetic field gradient of strength 6 G/cm in the $x$-$y$-plane, we remove the atoms in $m_F=+1$ and $m_F=-1$ and hence obtain a pure sample of atoms in $m_F=0$. Afterwards we ramp back the magnetic field to 47 G via a smooth S-ramp in 50 ms.

For the measurements with the spin mixture of $m_F=+1$ and $m_F=-1$, we start with atoms in $m_F=0$. As described above, we go smoothly to a small magnetic field of 114 mG and apply a resonant rf-pulse of 17 $\mathrm{\mu}$s which creates a state $\frac{1}{\sqrt{2}}(\ket{+1} + e^{i\phi_{2}}\ket{-1})$. We smoothly ramp back to a high magnetic field of 47 G before performing the measurements on the spin mixture. We achieve 49.6(1.5)\% in $m_F=+1$ and 50.4(1.4)\% in $m_F=-1$. Since all the measurements are performed in large magnetic fields, the relative phase $\phi_2$ between and $m_F=+1$ and $m_F=-1$ is not fixed in time.  

\begin{figure}
\includegraphics[width=8.7cm]{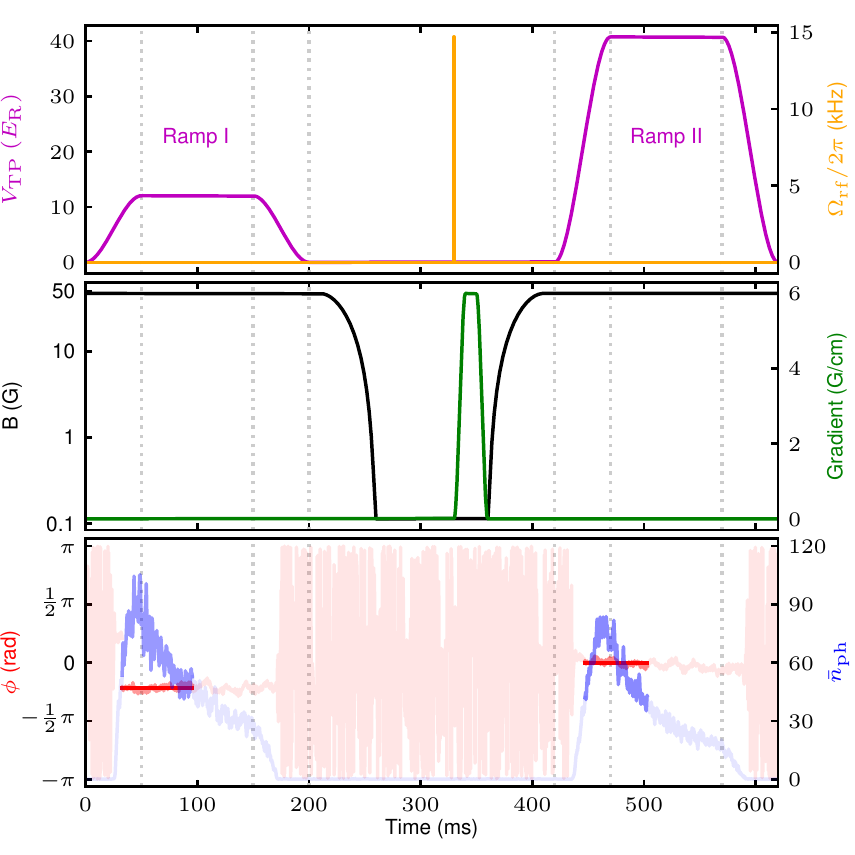}
\caption{Experimental ramps to detect the relative phase of the scattered light from different spin states during the process of self-organization. Top panel: Transverse pump lattice depth ramps (purple) used to perform self-organization. The radio frequency pulse (orange) is used to perform the spin flips. $\Omega_{\t{rf}}$ represents the Rabi frequency of coupling between different spin states via the radio frequency. Middle panel: Magnetic field ramp (black) and magnetic gradient ramp (green) which is used to remove spurious spin states in the preparation of $m_F=0$. Bottom panel: Exemplary trace of the self-organization process showing both the intracavity photon number $\bar{n}_{\t{ph}}$ (blue) and the phase $\phi$ of the detected light (red).}
\label{figSI}
\end{figure}

\subsection{Self-organization threshold measurement}
For the detection of the threshold lattice depth of individual spin states, we  ramp up the power of the transverse pump via an S-ramp in 100 ms. After a hold time of 3 ms, all the optical potentials are suddenly switched off. On average, we took 5 measurements per datapoint in Fig. 3a), with at least 4 and maximal 11 repetitions per data point.

For the construction of the phase diagram, we ramp up the power of the transverse pump via an S-ramp in 60 ms. After a hold time of 3 ms, we smoothly ramp down the transverse pump power in 50 ms and then switch off all the optical potentials suddenly. Fig. 2 shows the heterodyne signal averaged over 3 repetitions except for $m_F=0$ and $\varphi \simeq 90^{\circ}$, where only one repetition is shown.

For the threshold detection in a mixture of $m_F=\pm 1$, we ramp up the power of the transverse pump via an S-ramp in 50 ms. After a hold time of 20 ms, we smoothly ramp down the power of the transverse pump by applying an S-ramp of 50 ms duration and then suddenly switch off the remaining dipole potentials. On average, we took 3 measurements per datapoint in Fig. 4a), with at least 2 and maximal 6 repetitions per data point.

\subsection{Relative phase of the scattered light from different spin states}
We measure the relative phase of the scattered light from self-organization for different spin compositions via heterodyne detection of the scattered light by beating it with a local oscillator. In our experiment, the relative phase between the local oscillator and the transverse pump is stable within a few seconds but not from one experimental run to another. Hence we measure the relative phase between different spin states within a single experimental cycle \cite{Baumann2011as}.

For measuring the relative phase of the light scattered from single spin states, we start with all the atoms in $m_F=+1$. We ramp up the transverse pump power in 50 ms, hold for 100 ms duration to measure the phase of the organized state and then slowly ramp down the power in the transverse pump in 50 ms. We change the spin state to either $m_F=0$ or $m_F=-1$ and again ramp up the transverse pump power in 50 ms. After holding for 100 ms, we slowly ramp down the power of the transverse pump in 50 ms before switching off the remaining optical dipole potentials (Fig. \ref{figSI}). The time of flight pictures obtained in this way are further used to assess the quality of state preparation via a Stern-Gerlach separation. For the phase measurement where we flip the spin from $m_F=+1$ to $m_F=-1$, we perform post selection such that we achieve 82(10)\% of atoms in $m_F=-1$. On average, we took 5 measurements per datapoint in Fig. 3b), with at least 2 and maximal 11 repetitions per data point.

A similar procedure is repeated for the spin mixture while starting with $m_F=0$, that is, we measure the phase of the light scattered from the spin mixture with respect to the phase of the light scattered from $m_F=0$ state. The holding time in the organized phase is chosen to be 20 ms instead of 100 ms. On average, we took 5 measurements per datapoint in Fig. 4b), with at least 3 and maximal 10 repetitions per data point.

\section{Theory\label{Th}}
\subsection{Single multilevel atom coupled to the cavity mode}
The Hamiltonian of an atom coupled to the cavity field $\hat{H}_{\rm{atom-cav}}$ is given as
\begin{equation}
\hat{H}_{\rm{atom-cav}} = \hat{H}_{\rm{atom}} + \hat{H}_{\rm{cav}} + \hat{H}_{\rm{int}}.
\end{equation}

$\hat{H}_{\t{atom}}$ consists of the kinetic energy of the particle and the potential seen due to externally applied potentials as well as the internal energy of the atom
\begin{equation}
\hat{H}_{\t{atom}} = \frac{\hat{p}^2}{2m}+V_{\t{ext}}(\hat{\textbf{x}})+\sum_i E_i |i\rangle\langle i|,
\end{equation}
where the states $|i\rangle$ represents all possible atomic states. $\hat{H}_{\t{cav}}$ is the energy stored in the photon field of the cavity, in the rotating frame of the pump
\begin{equation}
\hat{H}_{\t{cav}} = -\hbar \Delta_{\t{c}} \hat{a}^\dagger \hat{a}.
\end{equation}
 The interaction operator $\hat{H}_{\t{int}}$ can be expressed, making use of the dipole and rotating wave approximation, as
\begin{equation}
\hat{H}_{\t{int}} = -\hat{\textbf{E}}^\dagger\cdot\hat{\textbf{d}}-\hat{\textbf{E}}\cdot\hat{\textbf{d}}^\dagger.
\end{equation}
Here $\hat{\textbf{E}}$ is the total electric field present at the position of the atom and $\hat{\textbf{d}}$ is the atomic dipole operator. The electric field can be decomposed into a classical contribution, coming from externally applied laser fields (pumps) and a quantum contribution from the cavity field 
\begin{equation}
\hat{\textbf{E}} =\boldsymbol{\epsilon}_{\t{p}} \frac{E_{\t{p}}}{2}  f(\hat{\textbf{x}}) e^{-i\omega_{\t{p}} t}+\boldsymbol{\epsilon}_{\t{c}}E_0 \hat{a}  g(\hat{\textbf{x}}) e^{-i\omega_{\t{p}} t},
\end{equation}
where we have restricted our description to elastic contributions, i.e. the cavity light is emitted at the pump frequency. $E_{\t{p}}$ represents the pump's electric field amplitude, $f(\hat{\textbf{x}})$ its spatial mode profile and $\boldsymbol{\epsilon}_{\t{p}}$ its polarization unit vector. $E_0$ is defined as in the main text. Finally $g(\hat{\textbf{x}})$ and $\boldsymbol{\epsilon}_{\t{c}}$ are the cavity's spatial mode profile and polarization, respectively.

We consider the case of dispersive coupling to the atoms and we restrict the description of the internal degrees of freedom to the ground state manifold, labeled by $|g\rangle$, all other levels will be indicated by $|e\rangle$. We treat the interaction term perturbatively at second order (the first order contribution is identically zero due to symmetry). The interaction Hamiltonian reduces to 
\begin{equation}
\hat{H}_{\t{int}}^{(2)} = \hat{\textbf{E}}^\dagger_i\hat{\alpha}_{i,j}\hat{\textbf{E}}_j
\end{equation}
 The Hermitian operator $\hat{\alpha}$ is a rank 2 tensor in the spatial indices and is given by
\begin{equation}
\hat{\alpha}_{i,j} = -\sum_{g,g'}\sum_e\frac{\langle g| \hat{d}_i|e\rangle\langle e|\hat{d}^\dagger_j|g'\rangle}{(E_e-E_g-\hbar\omega_{\t{p}})}|g\rangle\langle g'|.
\end{equation}
We can decompose the tensor $\hat{\alpha}$ into rank 0, rank 1, and rank 2 components $\hat{\alpha}_{{\t{s}},{\t{v}},{\t{t}}}$ \cite{LeKien2013s}, obtaining
\begin{eqnarray}
\hat{\alpha}_{\t{s}}&=&\alpha_{\t{s}} \hat{I}\delta_{ij},\\
\hat{\alpha}_{\t{v}}&=&-i\frac{\alpha_{\t{v}}}{2F} \epsilon_{ijk}\hat{F}_k,\\
\hat{\alpha}_{\t{t}}&=&\alpha_{\t{t}} \frac{3(\hat{F}_i\hat{F}_j+\hat{F}_j\hat{F}_i)-2F^2\hat{I}\delta_{ij}}{2F(2F-1)},
\end{eqnarray}
where we have introduced the total angular momentum operators for the ground state manifold $\hat{\textbf{F}}$, with maximum eigenvalue $F$. The values $\alpha_{{\t{s}},{\t{v}},{\t{t}}}$ depend solely on the electronic structure of the atom and the pump's frequency $\omega_{\t{p}}$. At the frequency used for the measurements presented in the main text, $\alpha_{\t{t}}$ is negligible for $^{87}$Rb. From now on we will restrict ourselves to linear polarization of the pump beam and cavity mode, i.e. $\boldsymbol{\epsilon}_{{\t{p}},{\t{c}}}$ are real. Without loss of generality we define $E_{\t{p}}$ as real, fixing the phase reference.  
Substituting, we find

\begin{equation}
\hat{H}_{\t{int}}^{(2)} = \frac{1}{4}\alpha_{\t{s}} f^2(\hat{\textbf{x}})E_{\t{p}}^2+\alpha_{\t{s}} g^2(\hat{\textbf{x}})E_0^2 \hat{a}^\dagger \hat{a}+\frac{1}{2}f(\hat{\textbf{x}})g(\hat{\textbf{x}})E_{\t{p}} E_0\left(\alpha_{\t{s}}\boldsymbol{\epsilon}_{\t{p}}\cdot\boldsymbol{\epsilon}_{\t{c}}(\hat{a}^\dagger+\hat{a})-i\frac{\alpha_{\t{v}}}{2F}\boldsymbol{\epsilon}_{\t{p}}\times\boldsymbol{\epsilon}_{\t{c}}\cdot\hat{\textbf{F}}(\hat{a}^\dagger-\hat{a})\right).\label{final}
\end{equation}

The first term gives rise to the optical potential generated by the pump, the second one can be visualized as a dispersive shift of the cavity resonance, the last term comes from the potential generated by the interference of the two fields and it gives rise to the physics explored in this manuscript. Taking a reference frame in which the external magnetic field points along $z$, $\hat{F}_x$ and $\hat{F}_y$ connect states differing by one unit of angular momentum, while the $z$-component is diagonal. If the magnetic field is large, $\mu_{\t{B}} B\gg\hbar\Delta_{\t{c}}$ as is the case in the experiment, $\hat{F}_x$ and $\hat{F}_y$ contributions can be neglected due to the high energy cost associated to the change of internal state. 

In the experiment, the mode functions are $f(\hat{\textbf{x}})=\cos(k \hat{z})$ and $g(\hat{\textbf{x}})=\cos(k \hat{x})$, where $k=2\pi/\lambda$. In this case the pump beam generates a lattice potential in the $z$-direction. The lattice amplitude is denoted as $V_{\t{TP}}$ and it is given by: $V_{\t{TP}}=-\alpha_{\t{s}} E_{\t{p}}^2/4$. We define the quantity $U_0=\alpha_{\t{s}} E_0^2/\hbar$ as the maximum shift in the cavity resonance caused by a single atom.  

\subsection{Many-body theory and two mode approximation for the BEC}
We focused so far on the description of the interaction of a single atom with the extended cavity mode. It is straight-forward to extend our analysis to the case of a BEC and cavity system. Using the formalism of second-quantization, we can write the many-body Hamiltonian as
\begin{equation}
\hat{H}_{\t{BEC}}=\hat{H}_{\t{cav}}+\int \hat{\Psi}^\dagger(\textbf{x}) \left(\hat{H}_{\t{atom}} + \hat{H}_{\t{int}}^{(2)}\right)\hat{\Psi}(\textbf{x}) d\textbf{x}
\end{equation} 
where $\hat{\Psi}$ is the many-body spinor wave function, normalized to the total atom number N. Here, we neglected direct atom-atom interactions in the low density limit. We will approximate $\hat{\Psi}$ as a product state. Following \cite{Baumann2010s}, we describe the BEC wave-function in momentum space. In the normal phase, the BEC occupies the ground state of the external potential. The potential is given by a weak harmonic trapping in all three spatial directions plus the pump's lattice potential in the $z$-direction. Neglecting the trapping potential, we can write the state of the BEC as 
\begin{equation}
|0\rangle=|k_x=0,k_y=0,q_z=0\rangle, 
\end{equation}
where the three entries represent momentum in $x$-and $y$-directions and quasi-momentum in the $z$-direction, due to the presence of the lattice potential generated by the transverse pump beam. 
The BEC can be prepared in each spin component. To span all possible states we use the set: $|0\rangle_{m_F}$, $m_F=-F,..,F$. We also introduce the set of states $|k\rangle_{m_F}$, where 
\begin{equation}
|k\rangle_{m_F}=\frac{1}{\sqrt{2}}\left(|\hbar k,0,\hbar k\rangle_{m_F}+|-\hbar k,0,\hbar k\rangle_{m_F}\right).
\end{equation}
  The interaction term couples the two states $|0\rangle_{m_F}$ and $|k\rangle_{m_F}$, with the same value of $m_F$. Reducing the description to (quasi-)momenta lower than $k$, we obtain the following form for the many-body Hamiltonian:

\begin{equation}
\hat{H}_{\t{BEC}}=-\left(\hbar \Delta_{\t{c}}-\frac{N U_0}{2}\right) \hat{a}^\dagger \hat{a}+\sum_{m_F}E(V_{\t{TP}})|k\rangle\langle k|_{m_F}
+M_0(V_{\t{TP}})E_{\t{p}} E_0\left[\alpha_{\t{s}}(\hat{a}^\dagger+\hat{a})\cos\varphi+i\frac{\alpha_{\t{v}}}{2F}(\hat{a}^\dagger-\hat{a})m_F\sin\varphi\right]\hat{J}_{x,m_F},
\end{equation}
 
where	$E(V_{\t{TP}})$ is the energy of state $|k\rangle$ and $M_0(V_{\t{TP}})$ is an overlap integral defined by $\langle 0|\cos(k x)\cos(k z)|k\rangle$.  These two quantities are modified due to the presence of the lattice and are functions of the lattice depth. We substituted $U_0\langle\cos^2(k x)\rangle=U_0N/2$ for the dispersive shift. This simplification gets rid of a non-linear term of the form $\hat{J}_z\hat{a}^\dagger\hat{a}$ in the final model. This term leads to interesting phenomenology, but does not influence the position of the critical point. The angle $\varphi$ is defined as in the main text. We define the operator $\hat{J}_{x,m_F}=\left(|0\rangle\langle k|_{m_F}+|k\rangle\langle 0|_{m_F}\right)/2$. 

Shifting the energy reference by $E(V_{\t{TP}})/2$, we obtain the form of the Hamiltonian reported in the main text

\begin{equation}
\hat{H}_{\t{BEC}}=-\hbar \widetilde{\Delta}_{\t{c}} \hat{a}^\dagger \hat{a}+\sum_{m_F}\left(\hbar\omega_0\hat{J}_{z,m_F}
+\frac{\hbar}{\sqrt{N_{m_F}}}\left(\lambda_{{\t{s}},m_F}(\hat{a}^\dagger+\hat{a})\cos\varphi+i\lambda_{{\t{v}},m_F}(\hat{a}^\dagger-\hat{a})m_F\sin\varphi\right)\hat{J}_{x,m_F}\right),\label{h}
\end{equation}

where we introduced the effective detuning $\widetilde{\Delta}_{\t{c}}=\Delta_{\t{c}}+\frac{N U_0}{2\hbar}$, the operators $\hat{J}_{z,m_F}=\left(|k\rangle\langle k|_{m_F}-|0\rangle\langle 0|_{m_F}\right)/2$, the energy of the atomic mode $\hbar\omega_0=E(V_{\t{TP}})$ and the coupling energies $\hbar\lambda_{{\t{s}},m_F}=M_0(V_{\t{TP}})E_{\t{p}} E_0\alpha_{\t{s}}\sqrt{N_{m_F}}$ and $\hbar\lambda_{{\t{v}},m_F}=M_0(V_{\t{TP}})E_{\t{p}} E_0\alpha_{\t{v}}\sqrt{N_{m_F}}/(2F)$. 

\subsection{Transition point for a single spin component}
In the case of a single spin component being present in the BEC, the Hamiltonian can be cast in the form of the Dicke model by a phase transformation on the $\hat{a}$ operator, $\hat{a}\rightarrow\hat{a}e^{-i\phi_{m_F}}$. The coupling energy for the resulting model becomes $\hbar|\lambda_{m_F}|$, where
\begin{eqnarray}
 &&|\lambda_{m_F}|=\sqrt{\lambda_{\t{s}}^2 \cos^2\varphi+\lambda_{\t{v}}^2 m_F^2\sin^2\varphi },\label{lc}\\
&&\phi_{m_F}=\arctan\left(\frac{\lambda_{\t{v}} m_F}{\lambda_{\t{s}}}\tan\varphi\right).\label{phi}
\end{eqnarray}
The transition point for the Dicke model in the presence of dissipation is given by $\lambda_{\t{c}}=\sqrt{(\widetilde{\Delta}_{\t{c}}^2+\kappa^2) \omega_0/(4\widetilde{\Delta}_{\t{c}})}$ \cite{Dimer2007s}. Using the expressions for $\lambda_{m_F}$ and $\lambda_{\t{c}}$, one obtains an expectation for the critical pump lattice depth as the solution of the equation
\begin{equation}
\frac{M_0^2(V_{\t{TP,c}})V_{\t{TP,c}}}{\hbar\omega_0(V_{\t{TP,c}})}=\frac{\widetilde{\Delta}_{\t{c}}^2+\kappa^2}{16\widetilde{\Delta}_{\t{c}} U_0 N_{m_F}\left(\cos^2\varphi+\left(\frac{\alpha_{\t{v}} m_F}{2F\alpha_{\t{s}}}\right)^2\sin^2\varphi\right)}.
\label{eq:th}
\end{equation}

\subsection{Phase boundaries for a spin mixture}
We now analyze the case of a balanced spin mixture in the $m_F=\pm 1$ states. In the following we will neglect the effects of cavity decay, these will be the subject of a future publication. We are going to analyze the system at the mean field level, defining $a=\langle\hat{a}\rangle$, $J_{i,\pm1}=\langle\hat{J}_{i,\pm1}\rangle$, where $i=x,y,z$. We start by performing adiabatic elimination of the cavity field, setting $\dot{a}=0$ in its equation of motion. The result of the elimination is
\begin{equation}
a=\frac{\lambda_1 J_{x,1}+\lambda_{-1}J_{x,-1}}{\sqrt{N/2}\widetilde{\Delta}_{\t{c}}}.
\end{equation}
If we plug this back in Eq. (\ref{h}), we obtain the mean field energy

\begin{equation}
\langle \hat{H}_{\t{BEC}}\rangle=-\hbar\omega_0\left(\sqrt{\frac{N^2}{16}-J_{x,1}^2-J_{y,1}^2}+\sqrt{\frac{N^2}{16}-J_{x,-1}^2-J_{y,-1}^2}\right)+\frac{\hbar|\lambda_1|^2}{N\widetilde{\Delta}_{\t{c}}}\left(J_{x,1}^2+J_{x,-1}^2+2\cos{2\phi_1}J_{x,1}J_{x,-1}\right),
\end{equation}

where we made use of conservation of particle number to eliminate the $J_z$ variables. We also set $N_1=N_{-1}=N/2$ and used $|\lambda_1|=|\lambda_{-1}|$ and $\phi_{1}=-\phi_{-1}$. We want to analyze the stability of the normal phase, defined by $J_{x,\pm1}=J_{y,\pm1}=0,J_{z,\pm1}=-N/4$, and therefore expand the energy around this point:

\begin{equation}
\langle \hat{H}_{\t{BEC}}\rangle \simeq \hbar\omega_0\left(\frac{2J_{x,1}^2}{N}+\frac{2J_{y,1}^2}{N}+\frac{2J_{x,-1}^2}{N}+\frac{2J_{y,-1}^2}{N}-\frac{N}{2}\right)+\frac{\hbar|\lambda_1|^2}{N\widetilde{\Delta}_\t{c}}\left(J_{x,1}^2+J_{x,-1}^2+2\cos{2\phi_1}J_{x,1}J_{x,-1}\right).
\end{equation}

We can see that a non zero value for $J_{y,\pm1}$ leads always to an increase in energy, therefore $J_{y,\pm1}=0$ in the ground state. We can get rid of the mixed term $J_{x,1}J_{x,-1}$ by a simple rotation, introducing the variables: $J_{x,{\t{d}}}=(J_{x,1}+J_{x,-1})/\sqrt{2}$ and $J_{x,{\t{s}}}=(J_{x,1}-J_{x,-1})/\sqrt{2}$. These will be the order parameters for density and spin organization, respectively.

\begin{equation}
\langle \hat{H}_{\t{BEC}}\rangle\simeq-\frac{\hbar\omega_0 N}{2}+\left(\frac{2\hbar\omega_0}{N}+\frac{\hbar|\lambda_1|^2}{N\widetilde{\Delta}_{\t{c}}}(1+\cos{2\phi_1})\right)J^2_{x,{\t{d}}}+\left(\frac{2\hbar\omega_0}{N}+\frac{\hbar|\lambda_1|^2}{N\widetilde{\Delta}_{\t{c}}}(1-\cos{2\phi_1})\right)J^2_{x,{\t{s}}}.
\end{equation}
 
For negative effective detuning, the coefficients of $J_{x,{\t{d(s)}}}^2$ can become negative, leading to an instability and a non zero value of the order parameter. For $\phi_1<\pi/4$, the coefficient of $J_{x,{\t{d}}}^2$ becomes negative for a lower value of $|\lambda_1|$. This means the system will enter the density organized phase from the normal phase. For $\phi_1>\pi/4$, the system enters instead the spin organized phase above the critical point. At the phase transition point the system breaks one of the two $Z_2$ symmetries of the Hamiltonian associated to the sign of $J_{x,{\t{d(s)}}}$. At $\phi=\pi/4$ the symmetry of the Hamiltonian becomes $Z_4$. This means that any direct transition between the density and spin organized phases will be of first order. From Eq. (\ref{phi}) we can evaluate the polarization angle corresponding to $\phi_1=\pi/4$, $\varphi_{\t{c}}=\arctan(2 F\alpha_{\t{s}}/\alpha_{\t{v}})$. Similarly, using Eq. (\ref{lc}) and the definition of $\lambda_{\t{s,v}}$, we obtain the expression for the critical lattice depth in the main text. 

\subsection{Accounting for collisional atomic interactions}

So far we restricted ourselves to the description of the system in absence of direct s-wave interactions between the atoms. These are nontheless present in our system and lead to a shift of the critical point for self organization. The behavior of the system is captured by the Gross-Pitaevski equation (GPE). It has been shown theoretically \cite{Nagy2008s} and verified experimentally \cite{Baumann2010s} that the main effect of interactions is to renormalize the value of the bare energy $\hbar\omega_0\rightarrow\hbar\omega_0+4E_{\t{int}}$ in the expression for the critical point with an interaction shift of the form:
\begin{equation}
E_{\t{int}}=\frac{g}{2N}\int|\psi_0(\textbf{x})|^4d\textbf{x},
\end{equation} 
where $g=4\pi\hbar^2 a/m$, $a$ is the scattering length and $\psi_0$ denotes the ground-state wave-function of the BEC as calculated from the solution of the GPE. To evaluate $\psi_0$, we employ a description based on a generalized local density approximation, as appropriate for our system due to the presence of harmonic trapping and a strong 1D lattice generated by the transverse pump. According to \cite{Kramer2003s}, we effectively partition the cloud into 'pancakes', the number of which is ranging from 8 to 16, depending on the value of $V_{\t{TP}}$. Along the $z$-direction, we utilize a non interacting form of the wave-function $\psi_z(z)=\langle z|q_z=0\rangle$, which is a good approximation for large lattice depths \cite{Kramer2003s}. For the radial density profile we employ the Thomas-Fermi approximation, resulting in
\begin{equation}
n_l(x,y)=\frac{1}{g_{\t{2d}}}(\mu-\frac{m}{2}\omega_z^2l^2d^2-\frac{m}{2}(\omega_x^2x^2+\omega_y^2x^2)),
\end{equation}
 where the interaction parameter was renormalized according to
\begin{equation}
g_{\t{2d}}=g \int{|\psi_z(z)|^4dz},
\end{equation}
$\omega_{x,y,z}$ are the trapping frequencies resulting from the trapping beams as well as the presence of the transverse pump lattice, $l$ is the pancake index and the chemical potential $\mu$ is chosen such that $\sum_l N_l=N$. The interaction energy is finally calculated according to
\begin{equation}
E_{\t{int}}=\frac{g_{\t{2d}}}{2N}\sum_l\int|n_l(x,y)|^2dxdy.
\label{eq:eint}
\end{equation}
\subsection{Interactions in the spin mixture}
For the case of a spin mixture in the $m_F=\pm1$ states, the interaction energy in the ground state reads 
\begin{equation}
E_{\t{int,mix}}=\frac{g_{0,\t{2d}}}{2N}\sum_l\int|n_{+1,l}(x,y)+n_{-1,l}(x,y)|^2dxdy+\frac{g_{1,\t{2d}}}{2N}\sum_l\int|n_{+1,l}(x,y)-n_{-1,l}(x,y)|^2dxdy, 
\label{eq:eint,spin}
\end{equation}
where $n_{m_F}$ is the density for each spin component and $g_{0(1),\t{2d}}$ is the spin independent (dependent) interaction parameter obtained by replacing the scattering length with $a=2a_2+a_0$ $(a=a_2-a_0)$ respectively, in the expression for $g_{\t{2d}}$ \cite{Stamper-Kurn2013s}. For $^{87}$Rb in the F=1 manifold, $(2a_2+a_0)/(a_2-a_0)\simeq200$, so that we can neglect the second term in Eq. (\ref{eq:eint,spin}).
Following \cite{Nagy2008s}, the bare energy of the excited state is renormalized according to $\hbar\omega_0\rightarrow\hbar\omega_0+4E_{\t{int,mix}}$ for organization in the density modulated state. This is not the case though when considering organization in the spin texture, because atomic density is only weakly modified when entering this phase. No renormalization of $\omega_0$ is therefore expected. 

So far we assumed that the different spin components of the BEC have the same spatial mode. This might not be the case in the experiment due to the presence of spurious magnetic field gradients that can partially separate the clouds. From the form of Eq. (\ref{eq:eint,spin}) one can easily verify that the interaction energy for two spatially separated clouds is a factor of 2 lower than for perfectly overlapping ones. Depending on the overlap, the interaction energy $E_{\t{int,mix}}$ can therefore range between $E_{\t{int}}$ and $E_{\t{int}}/2$. If the clouds are separated, the atomic density is modified in exactly the same way when entering the phase transition for both the density modulated phase and the spin texture. In this case the shift of $\hbar\omega_0$ is therefore $2E_{\t{int}}$ for both phases.

\section{Data Processing}
\subsection{Cavity output detection}
The cavity has two linearly polarized TEM$_{00}$ eigenmodes which are lying in the $y$-$z$-plane and are rotated by $\alpha=22^{\circ}$ with respect to the $y$- respectively $z$-axis. The birefringence of the cavity separates the two eigenmodes by $\delta_{\t{B}}/2\pi=2.2$ MHz. The setup is aligned such that the $y$-polarized component of the light leaking from the cavity is detected by a heterodyne setup while the $z$-polarized component is detected by a single photon counting module (SPCM). The heterodyne detector is used to extract information about the number of intracavity photons and the phase of this light, while the SPCM is used to probe the cavity resonance close to $z$-polarization after every experimental cycle. An averaging window of 1 ms is used to process both heterodyne and SPCM data.\\

\subsection{Self-organization threshold detection and phase detection}
\paragraph{Obtaining the thresholds for self-organization.} For every data trace, we fit the initial rise in the photon number extracted from the heterodyne detector (up to $50\%$ of the maximum signal) with a piecewise linear and power law function as a function of lattice depth $V_{\t{TP}}$ to determine the threshold lattice depth $V_{\t{TP}}^{\t{c}}$ for self-organization \cite{Landig2016s}. The error bars are the sum of the statistical and the systematic errors. The latter is obtained from the variation in threshold obtained by changing the range where the time traces are fitted from 30\% to 70\% of the maximum intra-cavity photon number. This systematic error amounts to 5\% of the mean value $V_{\t{TP}}^{\t{c}}$.\\

\paragraph{Obtaining the phase data}
We evaluate the mean value of the phase of the scattered light during a self-organization process by using the fraction of data where the intra-cavity photon number is at least 50\% of the maximum value (Fig. \ref{figSI}). In this way, we obtain the phase difference $\phi$ of light scattered from (potentially) different spin configurations in the same experimental run by subtracting the corresponding mean values. The error bars are the sum of the statistical error and the systematic error. The systematic error $\delta \phi ^{\t{syst}}$ is estimated by analyzing the relative phase $\phi$ while organizing twice in the $m_F = +1$ ($m_F = 0$) spin state for the case of single spins (the spin mixture). The spread of these values, measured with the standard deviation, gives the systematic uncertainty $\delta \phi ^{\t{syst}}$ in the phase. This systematic shift, which should ideally be zero, is mainly due to the drift in the relative phase of the local oscillator and the transverse pump and amounts to $\delta \phi ^{\t{syst}}\simeq 5.5^{\circ}$.

\subsection{Phase diagram}
In order to plot the phase diagram in Fig. 2, the data from both SPCM and heterodyne detector are converted to number of photons and summed up. This is crucial when the frequency difference between the cavity and transverse pump is small which makes the role of both cavity birefringent modes important. The detuning $\Delta_{\t{c}}^{\prime} = \omega_{\t{p}} - \bar{\omega}_{\t{c}}$ in Fig. 2 is defined using the frequency $\bar{\omega}_{\t{c}}$ which is the mean of the resonance frequencies of the two cavity eigenmodes. \\

\subsection{Theoretical lines / fits}
\paragraph{Fitting the phase data.} 
In order to calibrate the angle $\varphi$ of the linear polarization of the transverse pump, we place a polarization analyzer at the exit of the vacuum chamber. Due to physical constraints of the experimental apparatus, it is challenging to align the axis of the polarization analyzer with the orientation of the cavity inside the vacuum chamber. This misalignment gives rise to an angle offset $\varphi_0$, such that the polarization analyzer therefore reads the quantity $\varphi'=\varphi-\varphi_0$. We estimate the offset to be $\approx 70^{\circ}$. In order to calibrate the angle offset $\varphi_0$ more reliably, we perform a common fit for the two datasets in Fig. 3b to determine $\varphi_0$.
The phase of the light leaking out of the cavity in the self-organized phase as a function of polarization angle $\varphi=\varphi'+\varphi_0$ is depending on the initial and final spin state and is described by
\begin{align}
\phi_{1\rightarrow 0}(\varphi',\varphi_0)&=\phi_{m_F=0}-\phi_{m_F=1}=-\arctan{\left(\frac{\alpha_{\t{v}}}{2\alpha_{\t{s}}} \tan{(\varphi'+\varphi_0)}\right)},\\
\phi_{1\rightarrow -1}(\varphi',\varphi_0)&=\phi_{m_F=-1}-\phi_{m_F=1}=-2 \arctan{\left(\frac{\alpha_{\t{v}}}{2\alpha_{\t{s}}} \tan{(\varphi'+\varphi_0)}\right)},
\end{align}
where $\frac{\alpha_{\t{v}}}{2\alpha_{\t{s}}}=0.464$ is fixed.
 We checked that the imperfect atom number preparation for $m_F=-1$ has a negligible effect on the value of $\varphi_0$ by estimating it using only the dataset $\phi_{1\rightarrow 0}$. We subtract the fitted curve for the dataset $\phi_{1\rightarrow 0}$ from every data point to symmetrize the phases around $m_F=0$. The symmetrized data for the final spin states $m_F=\pm 1$ are fitted again with a common $\varphi_0$ 
\begin{align}
\phi_1(\varphi',\varphi_0)&=+\arctan{\left(\frac{\alpha_{\t{v}}}{2\alpha_{\t{s}}} \tan{(\varphi'+\varphi_0)}\right)},\\
\phi_{-1}(\varphi',\varphi_0)&=- \arctan{\left(\frac{\alpha_{\t{v}}}{2\alpha_{\t{s}}} \tan{(\varphi'+\varphi_0)}\right)}.\\
\end{align}
This gives the angle offset $\varphi_0=66.9(1.1)^{\circ}$, which is used for all data presented in this paper. 
\paragraph{Corrections from transverse pump lattice and collisional atomic interactions.} In order to incorporate corrections to our measured threshold lattice depths $V_{\t{TP}}^{\t{c}}$ from the presence of the transverse pump lattice and collisional atomic interactions, as described in Eq. \eqref{eq:th} and \eqref{eq:eint}, we map each $V_{\t{TP}}^{\t{c}}$ via the function $h$ onto the corrected value 
\begin{align}
h_{\beta}: V_{\t{TP}}^{\t{c}} \rightarrow V_{\t{TP,corr}}^{\t{c}}=\frac{E_{\t{R}} M_0^2(V_{\t{TP}}^{\t{c}})}{\hbar \omega_0(V_{\t{TP}}^{\t{c}})+ \beta E_{\t{int}}(V_{\t{TP}}^{\t{c}})}V_{\t{TP}}^{\t{c}}.
\end{align}
The parameter $\beta$ is set to account for different interaction strengths present in different self-organization regimes and imperfect overlap between the clouds in the spin mixture measurement.
We fit the corrected threshold lattice depths $h_{\beta}(V_{\t{TP}}^{\t{c}})$. For depiction in the figures, we show the measured threshold lattice depths $V_{\t{TP}}^{\t{c}}$ and $h^{-1}_{\beta}(V_{{\t{TP,corr}}}^{\t{c}})$.
\paragraph{Threshold lattice depth for single spin clouds.} 
According to Eq. \eqref{eq:th}, we describe the threshold lattice depth for a BEC containing atoms of only one spin state by
\begin{align}
V_{{\t{TP}},m_F}^{\t{c}}(\varphi,c)=\frac{c}{\cos^2{(\varphi)}+(\frac{\alpha_{\t{v}} m_F}{2F\alpha_{\t{s}}})^2 \sin^2{(\varphi)} },
\end{align}
where $c$ is a global scaling factor, and $\frac{\alpha_{\t{v}}}{2\alpha_{\t{s}}}=0.464$ is fixed. We perform a common fit over all three datasets for $m_F=0,\pm 1$ on $h_{\beta=4}(V_{\t{TP},m_F}^{\t{c}})$, obtaining a global scaling factor of $c=0.519(4)$~$E_{\t{R}}$. 
We plot $h^{-1}_{\beta=4}(V_{{\t{TP}},m_F}^{\t{c}})$ in Fig. 3(a).
\paragraph{Self-organization threshold lattice depth for mixture of $m_F=\pm 1$.}
The density organization threshold for a gas containing an equal number of atoms with $m_F=\pm 1$ spin is described by
\begin{align}
V_{{\t{TP,mix}}}^{\t{c,d}}(\varphi)&=\frac{c_{\t{mix}}}{\cos^2{(\varphi)}+(\frac{\alpha_{\t{v}} m_F}{2F\alpha_{\t{s}}})^2 \sin^2{(\varphi)}}\frac{2}{+\cos{(2 \arctan{(\frac{\alpha_{\t{v}} m_F}{2F\alpha_{\t{s}}} \tan{(\varphi)})})}+1},
\label{eq:thmixdens}
\end{align}
while the threshold for spin organization is described by 
\begin{align}
V_{{\t{TP,mix}}}^{\t{c,s}}(\varphi)&=\frac{c_{\t{mix}}}{\cos^2{(\varphi)}+(\frac{\alpha_{\t{v}} m_F}{2F\alpha_{\t{s}}})^2 \sin^2{(\varphi)}}\frac{2}{-\cos{(2 \arctan{(\frac{\alpha_{\t{v}} m_F}{2F\alpha_{\t{s}}} \tan{(\varphi)})})}+1}.
\label{eq:thmixspin}
\end{align}
where $c_{\t{mix}}=0.48(6)$~$E_{\t{R}}$ is a global scaling factor, obtained from scaling $c$ from the single spin measurements by the atom number, and $\frac{\alpha_{\t{v}}}{2\alpha_{\t{s}}}=0.464$ is fixed as before.\\
The shaded area in Fig. 4(b) of the main text depicts the region where Eq. \eqref{eq:thmixdens} and \eqref{eq:thmixspin} predict the threshold lattice depths, incorporating the effect of different strengths of collisional interactions in the density- or magnetization modulated regime as well as imperfect overlap between the different clouds. The lower (higher) boundary of the threshold curve which is increasing for increasing polarization angle $\varphi$, therefore describing density organization, originates from assuming no (full) overlap between the $m_F=\pm 1$ clouds and thus using $h^{-1}_{\beta=2}(V_{\t{TP,mix}}^{\t{c,d}})$ ($h^{-1}_{\beta=4}(V_{\t{TP,mix}}^{\t{c,d}})$). Accordingly, the lower (higher) boundary of the threshold curve describing spin organization is calculated by using $h^{-1}_{\beta=0}(V_{\t{TP,mix}}^{\t{c,s}})$ ($h^{-1}_{\beta=2}(V_{\t{TP,mix}}^{\t{c,s}})$).

\bibliographystyle{apsrev4-1}
\end{widetext}
\end{document}